\documentclass[twocolumn]{aastex61}

\usepackage{amsmath, amsthm, amssymb}
\usepackage{graphicx}
\usepackage{arydshln}
\usepackage{bm}
\usepackage{xcolor}
\usepackage{epstopdf} 
\usepackage{colortbl} 
\usepackage[utf8x]{inputenc}
\accepted{\today}
\submitjournal{ApJ}

\shorttitle{Mixing in giant impacts}
\shortauthors{Deng et, al.}

\begin{document}

\title{\emph{Enhanced} mixing in giant impact simulations with a new Lagrangian method}

\correspondingauthor{Hongping Deng}
\email{hpdeng@physik.uzh.ch}

\author{Hongping Deng}
\affiliation{Center for Theoretical Astrophysics and Cosmology, Institute for Computational Science, University of Zurich, Winterthurerstrasse 190, 8057 Zurich, Switzerland}

\author{Christian Reinhardt}
\affiliation{Center for Theoretical Astrophysics and Cosmology, Institute for Computational Science, University of Zurich, Winterthurerstrasse 190, 8057 Zurich, Switzerland}

\author{Federico Benitez}
\affiliation{Center for Theoretical Astrophysics and Cosmology, Institute for Computational Science, University of Zurich, Winterthurerstrasse 190, 8057 Zurich, Switzerland}

\author{Lucio Mayer}
\affiliation{Center for Theoretical Astrophysics and Cosmology, Institute for Computational Science, University of Zurich, Winterthurerstrasse 190, 8057 Zurich, Switzerland}

\author{Joachim Stadel}
\affiliation{Center for Theoretical Astrophysics and Cosmology, Institute for Computational Science, University of Zurich, Winterthurerstrasse 190, 8057 Zurich, Switzerland}

\author{Amy C. Barr}
\affiliation{Planetary Science Institute, 1700 E. Ft. Lowell, Suite 106, Tucson, AZ 85719, USA}
\begin{abstract}
Giant impacts (GIs) are common in the late stage of planet formation. The Smoothed Particle Hydrodynamics 
(SPH) method is widely used for simulating the outcome of such violent collisions, one prominent example 
being the formation of the Moon. However, a decade of numerical studies in various areas of 
computational astrophysics has shown that the standard formulation of SPH suffers from several 
shortcomings  such as artificial surface tension and its tendency to promptly damp turbulent motions on scales much larger
than the physical dissipation  scale, both resulting in the suppression of mixing.
In order to estimate how severe these limitations are when
modeling GIs we carried out a comparison of simulations with identical initial conditions performed with
the standard SPH as well as with the novel Lagrangian Meshless Finite Mass (MFM) method using the multi-method code, GIZMO \citep{Hopkins2015}. 
We confirm the lack of mixing between the impactor and target when SPH is employed, while
MFM is capable of driving vigorous subsonic turbulence and leads to significant mixing between the two bodies. Modern
SPH variants with artificial conductivity, a different formulation of the hydro force or 
reduced artificial viscosity, do not improve mixing as significantly. Angular momentum is conserved similarly
well in both methods, but MFM does not suffer from spurious transport induced by artificial viscosity, resulting
in a slightly higher angular momentum of the protolunar disk. Furthermore, SPH initial conditions 
unphysically smooth the core-mantle boundary which is easily avoided in MFM.
\end{abstract}

\keywords{Giant impact, fluid mixing, numeric-Lagrangian method}



\section{Introduction} \label{sec:intro}
During the late stage of terrestrial planet formation, energetic collisions between roughly Mars-sized planetary embryos are common \citep{Chambers2001}. These collisions are called giant impacts (GIs) and influence the mass, spin, and the number of planets in the final planetary system. The outcome of such violent collisions have been studied in many previous publications \citep{Asphaug2006, Leinhardt2012}. One particularly compelling case is the giant impact hypothesis for the formation of the Moon \citep{Cameron1976, Benz1986, Canup2001}. The Moon and the Earth have almost identical isotope composition for several elements, such as oxygen \citep{Wiechert2001} and titanium \citep{Zhang2012}. Either the impactor has very similar isotopic composition to the proto-Earth \citep{Dauphas2017,Mastrobuono2015} or the impact mixes them efficiently assuming every planetary mass body has a unique isotopic signature \citep{Kaib2015}\citep[see review by][]{Barr2016}. Many Smoothed Particle Hydrodynamics (SPH) simulations found that most disk silicates are derived from the impactor and mixing seems insufficient \citep{Canup2013}. Alternative models like a fast-spinning proto-Earth \citep{Cuk2012}, a hit and run collision \citep{Reufer2012} and an impact between bodies of roughly equal mass \citep{Canup2012} have been proposed. However, all models are not entirely satisfactory because they either fail to explain the observations or introduce new issues, for example, forming a fast-spinning proto-Earth, which need to be solved.

Most GI simulations have used SPH \citep{Lucy1977, Gingold1977}. A few Eulerian code simulations are available, such as with the FLASH code \citep{Fryxell2000,Liu2015} and the CTH code \citep{CTH1990, Canup2013}. Many shortcomings of SPH have been exposed and overcome in the past few years, such as the artificial tension force acting at the interface between two fluids \citep{Agertz2007, Price2008}, 
and the excessive numerical viscosity in shear flow \citep{Cullen2010}.
A new SPH formulation has been proposed \citep{Saitoh2013,Hopkins2013} and used in GI simulations by \citet{Hosono2016}. Special techniques for SPH are also developed in GI simulations, such as the treatment of free surface and the explicit conservation of entropy \citep{Reinhardt2017}. 
Discreteness particle noise in SPH and artificial viscosity smear out local velocity variations thus damping
subsonic turbulence on overly large scales relative to the physical dissipation scales of the turbulent cascade \citep{Bauer2012}. These issues have promoted improvements of
the method \citep{Beck2016} which are absent in all previous giant impact simulations using SPH. Alternatively, other hydrodynamical solvers have recently been developed that still keep the main advantage of SPH in treating collisions between bodies,
namely its Lagrangian nature.
\cite{Hopkins2015} implemented a new Lagrangian meshless finite mass (MFM) method in the GIZMO code showing excellent shock capturing and 
conservation properties \citep{Hopkins2015, Deng2017}. \citet{Hopkins2015} also shows that MFM can capture  small-scale turbulence,
yielding results that are very similar to those of moving-mesh and stationary-grid methods. GIZMO MFM also appears to sustain subsonic MRI 
\citep{Balbus1991} turbulence much longer than SPH in local shearing box simulations (Deng et al. 2018, in prep).

We ran GI simulations using the multi-method GIZMO code \citep{Hopkins2015}, employing both MFM and SPH for different equations of state and planetary compositions to investigate the role of the numerical hydrodynamics method on mixing in the post-impact target. We also analysed the protolunar disk's dynamic property and composition.
The main features of the hydrodynamical methods adopted and the initial conditions of GIs
are described in section~\ref{sec:methods}. We present the results of single component impacts in section~\ref{sec:single} 
as well as multiple component impacts in section~\ref{sec:multiple}. 
We discuss the results in section~\ref{sec:discuss} and draw conclusions in section~\ref{sec:conclusions}.

\section{Numeric Methods}
\label{sec:methods}
\subsection{The hydro-methods}
 We use the GIZMO code \citep{Hopkins2015} which includes a number of particle-based hydro solvers, and have augmented them with new equations of state in order
to be able to model giant impact (GIs). In particular, we use the standard SPH solver inherited from the
GADGET3 code (see \citet{Springel2005}) which is based on the density-energy formulation of the SPH equations and adopts standard Monaghan artificial
viscosity with the Balsara switch \citep{Balsara1995} to minimize viscous dissipation away from shocks. The other numerical hydrodynamics method that
we consider is MFM, which solves the hydro equations by partitioning the domain using volume elements associated with the original
particle distribution, and computing fluxes at the interfaces of the resulting tessellation by means of a Riemann solver as in finite volume Godunov-type methods\citep{Hopkins2015}. While many modern SPH variants have appeared in the last years that improve considerably in its ability to model complex flows, we
chose to use this relatively old SPH formulation to enable comparison with most past work. However, we tested the effect of improvements present in modern SPH codes such as the 
Cullen \& Dehnen artificial viscosity switch \citep{Cullen2010} and the artificial thermal conductivity of \citet{Read2012} in the discussion section(see section \ref{sec:discuss}). \citet{Hosono2016} presented GI simulations with density independent SPH (DISPH) \citep{Saitoh2013,Hopkins2013}. However, it is not trivial to enable non-ideal equation of state (EOS) in DISPH \citep{Hosono2013}. We present no DISPH simulations since DISPH also damps subsonic turbulence (our focus of the paper, see figure \ref{fig:vel},\ref{fig:vel2}) as SPH \citep{Hopkins2015}.

The newest version of GIZMO \footnote{The public version of the code, containing all the algorithms used here, is available at \href{http://www.tapir.caltech.edu/~phopkins/Site/GIZMO.html}{\url{http://www.tapir.caltech.edu/~phopkins/Site/GIZMO.html}}} \citep{Hopkins2017} supports a general EOS (including Tillotson EOS interface) implemented by the author of the code. We added in our own EOS interface. The HLLC (Harten-Lax-van Leer-Contact) Riemann solver \citep{Toro1994} is extended for general EOS by doing explicit state reconstruction for the sound speed and internal energy.  The Riemann solver works well with general EOS, see appendix \ref{sec:test1}. In order so assess numerical issues due to this generalized Riemann solver we also tested a more accurate contact wave estimation proposed by \citet{Hu2009} but find that there is no noticeable difference to the default HLLC solver so we did not use it in the simulations presented in this paper.

\begin{figure}[ht!]
\plotone{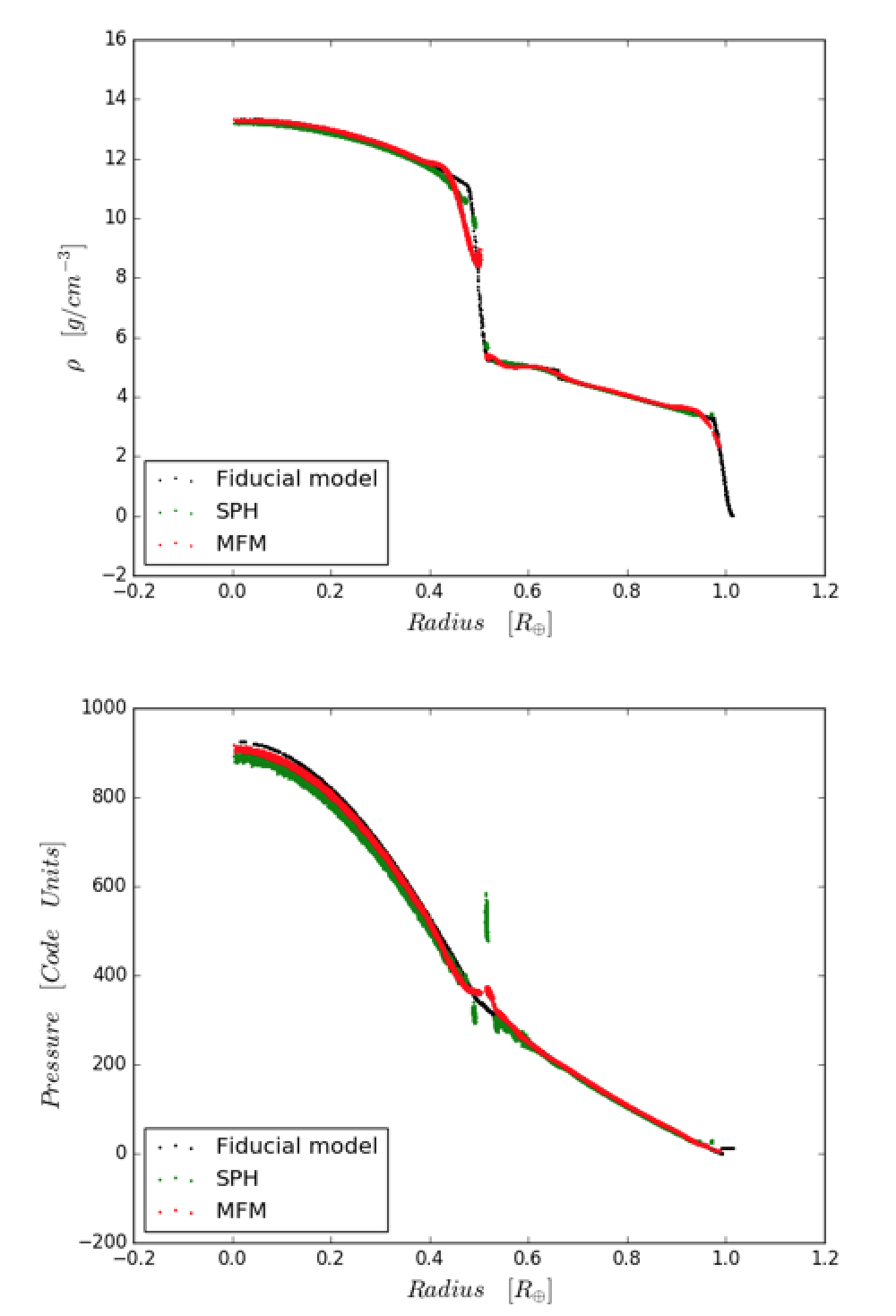}
\caption{The density (upper panel) and pressure (lower panel) profile of the $0.89M_{\oplus}$ target in the \emph{benchmark} moon formation run119 of \citet{Barr2016}. The initial condition is modeled with 500K particles of equal mass. The CTH grid code model (fiducial model), SPH model and MFM model are shown in black, green and red respectively. Some particles/cells enter an unphysical state in the core-mantle transition region in all three models with the SPH model showing a non-continuous pressure profile at the core-mantle boundary. \label{fig:eqm}}
\end{figure}

We use the Tillotson equation of state (EOS) \citep{Tillotson1962} to model impacts of undifferentiated objects and ANEOS/M-ANEOS \citep{Thompson1974, Melosh2007} for a multiple-component impact model (a differentiated structure with 30\% iron (ANEOS) and 70\% dunite (M-ANEOS) by mass). The Tillotson EOS does not yield a thermodynamically consistent treatment of mixtures between two phases, and can not model the critical behaviour at phase transitions \citep{Brundage2013}. However, pressure-release melting might happen when the highly compressed core is unloaded from equilibrium \citep{Asphaug2006}. We show that the ANEOS EOS does capture the phase transition in the iron core when it is strongly disturbed in appendix \ref{sec:test2}. We use 1 Earth radius ($R_{\oplus})$, 1 km/s with the gravitational constant equals 1 as our unit system. We describe our core-mantle boundary treatment in the following section.

\begin{figure}[ht!]
\plotone{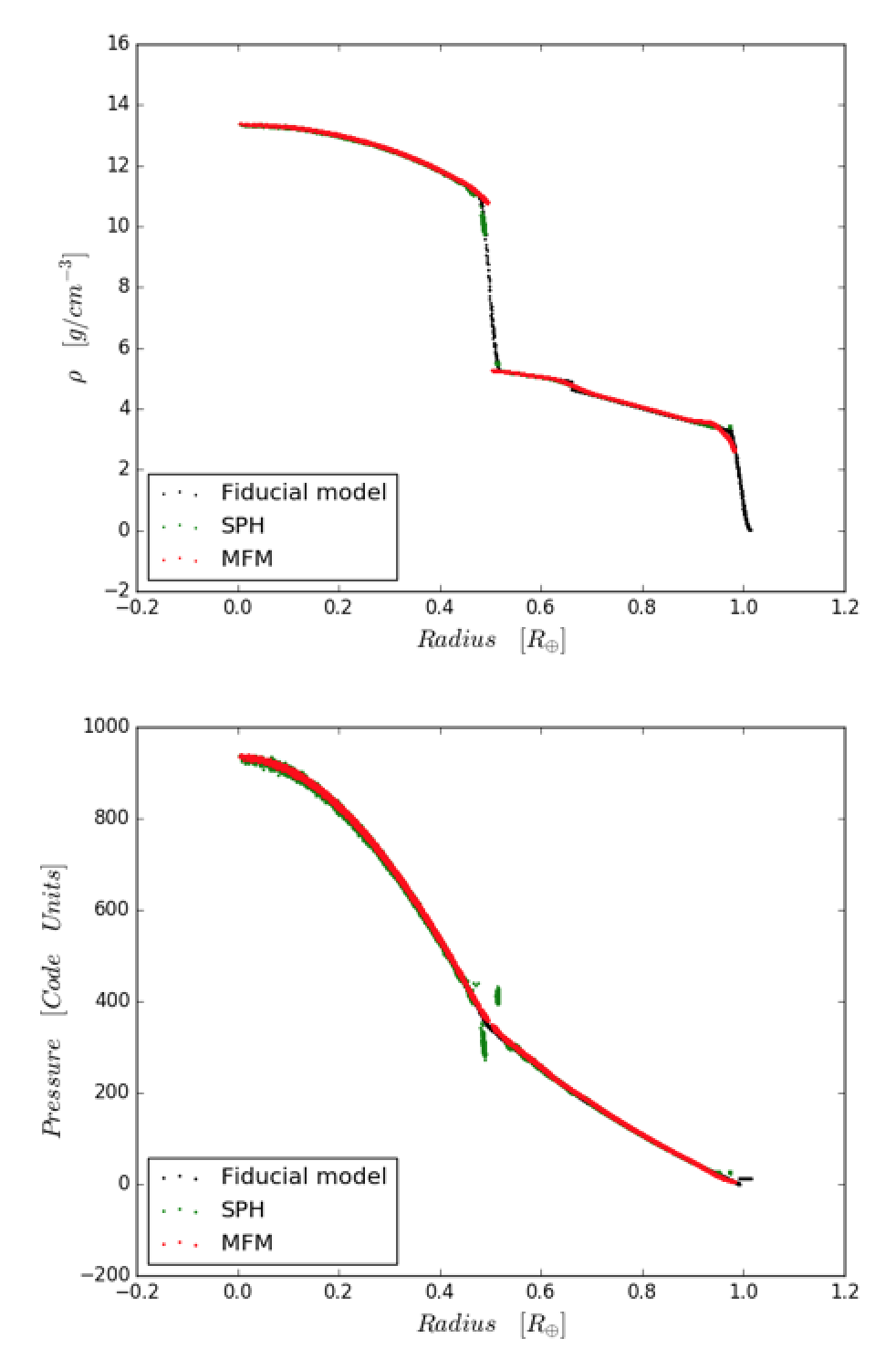}
\caption{The density (upper panel) and pressure (lower panel) profile of the $0.89M_{\oplus}$ target in the \emph{benchmark} Moon formation run119 of \citet{Barr2016}. The initial condition is modeled using 500K particles with iron particles' mass equal two times dunite particles' mass. The CTH grid code model (fiducial model), SPH model and MFM model are shown in black, green and red respectively. Only the MFM model kepng the infinitely sharp (no low-density iron particles) core-mantle transition while the SPH model still has non-continuous pressure profile at the core-mantle boundary. \label{fig:2m}}
\end{figure}

\subsection{Initial conditions and the Core-mantle boundary}
\label{sec:coremantle}
We follow \citet{Reinhardt2017} to produce a low noise representation of a planet's equilibrium initial conditions based on equal area tessellations of the sphere. The initial setups are further relaxed by running them with the hydro code chosen for the run (standard SPH or MFM)
for about 3 hours of simulation time  until the random velocity of particles, measured by their root mean square velocity, 
is less than 1\% of the impact velocity. 
In order to avoid problems at the planet's surface while relaxing the model, we applied the free surface treatment proposed in \citet{Reinhardt2017} but disabled it during the impact simulation to allow a direct comparison with published results. Removing the free surface treatment has no effects on the planet’s thermal state on the short timescale of the initial collision, except in the very outer part.

We use 500K particles (comparable to recent high-resolution impact simulations) to sample the target ($0.89 M_{\oplus}$) in the canonical Moon formation scenario \citep{Canup2013}. It is isentropic with an entropy of 1200J/kg/K in the core and 2700J/kg/K in the mantle \citep[see][for details]{Barr2016}. In SPH, the density of the $i$th particle is the kernel weighted sum of its neighbor particles' masses \citep{Springel2005};

\begin{equation}
    \rho_i=\Sigma_j m_j W(\vert\bm{r}_i-\bm{r}_j\vert,h_i),
\end{equation}

as a result, the core-mantle boundary is not infinitely sharp. The core-mantle transition 
is at the smoothing length scale in SPH while MFM has a larger transition region 
(see the upper panel of figure \ref{fig:eqm}). Particles/cells in 
the transition region with a density 
intermediate between that of iron and dunite do not have well defined physical properties.
They are \emph{expanded iron} or \emph{compressed dunite} in the EOS table which is not physically motivated.

\begin{figure*}[ht!]
\plotone{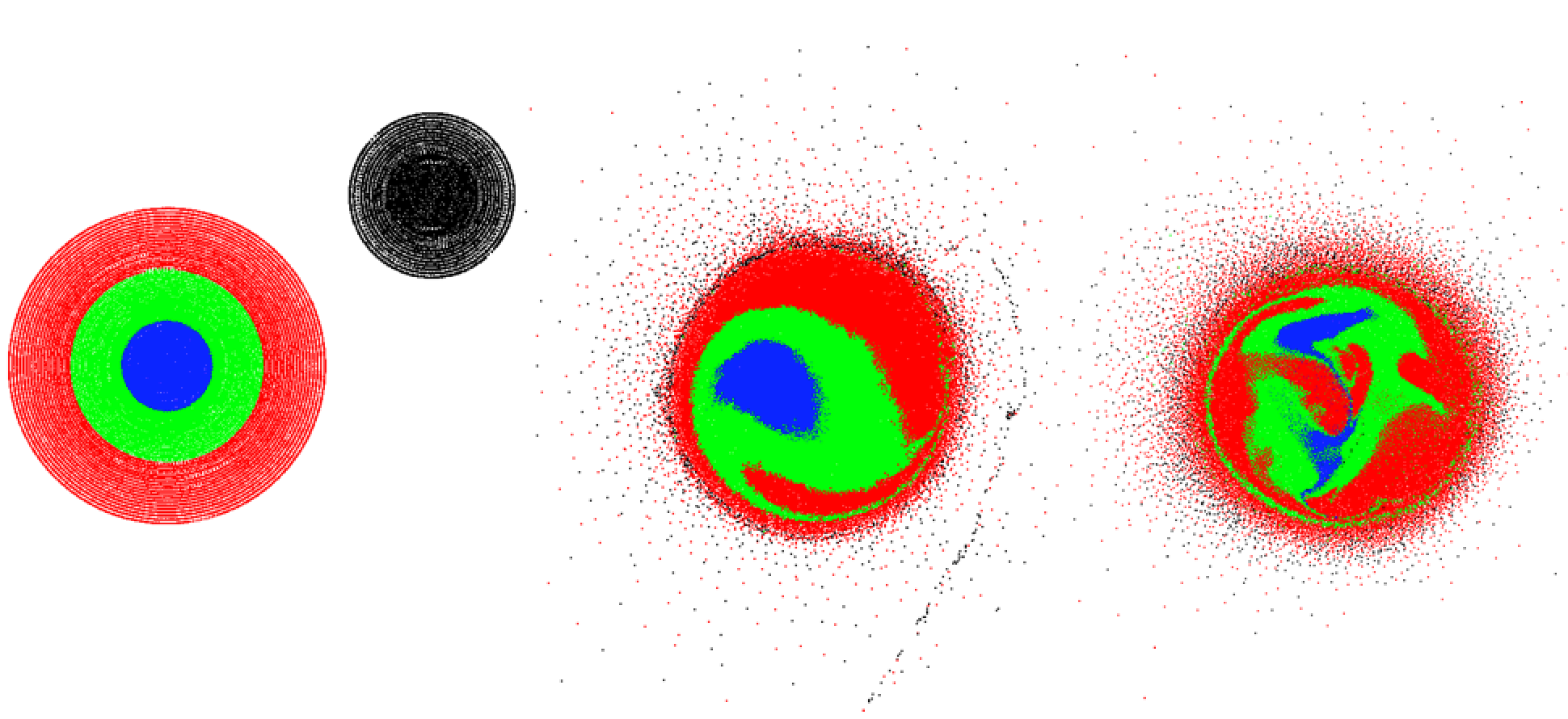}
\caption{Single componenet (Tillotson granite) impact. The left panel shows color-labeled different layers (slice between $-0.1<z<0.1$) of the pre-impact target and the impactor. The inner structure of the post-impact target (slice between $-0.1<z<0.1$ ) at $t=13.8\,h$ are  shown in the middle (run with SPH) and right (run with MFM) panel. The center is disrupted and even some particles from the impactor get into the innermost region in the MFM simulation while the SPH simulation only show moderate deformation of the target.\label{fig:ids}}
\end{figure*}
Additionally, at the core-mantle boundary, the density, and thus the smoothing length, changes sharply.  This leads to an artificial tension force separating the two components in standard SPH \citep{Agertz2007, 
Price2008}. In the lower panel of figure \ref{fig:eqm}, for the SPH realization, 
we notice a discontinuous pressure profile when employing the M-ANEOS EOS. This is 
caused by artificial surface tension. Instead,  
MFM delivers a continuous pressure profile, albeit still exhibiting
a small pressure bump.  Surface tension prevents fluid mixing  \citep{Agertz2007},
but preserves a sharper core-mantle boundary in standard SPH compared to MFM (see figure \ref{fig:eqm}).
 
 \citet{Woolfson2007} proposed an extra correction factor for the density at the interface 
between different components to maintain a sharp core-mantle transition. However, this
is an {\it ad hoc} correction which is not formally consistent with the 
SPH or MFM formulation. We follow a different strategy and use particles with different masses 
in our MFM model.  We recall that, in MFM, the density of the $i$th particle is:
\begin{equation}\label{eq:rho}
  \rho_{i} = \frac{m_{i}}{V_{eff,i}} ,
\end{equation}
where $V_{eff,i}$ is the effective volume of the $i$th particle  \citep[see][]{Hopkins2015}. 
Using iron particles of mass two times that of the dunite particles', 
the smoothing length, thus $V_{eff}$, is almost continuous across the core-mantle boundary, yet 
we obtain sharp core-mantle boundary with no particles entering an unphysical state 
(see upper panel of Figure \ref{fig:2m}). \citet{Woolfson2007} 
had to vary the correction factor according to the density ratio of the two components whereas with our approach
we simply use a 2:1 mass ratio of particles. Indeed moderate variations in the density ratio 
are tolerable while a time-dependent variation of particle mass would cause the method to fail. 
In figure \ref{fig:2m}, the pressure is still continuous in the MFM model, and overlaps with the 
fiducial model, while the SPH model still suffers from artificial tension force 
and has particles entering unphysical states. 
In the impact simulations, we use different mass particles in MFM but the same mass particles 
for SPH to enable direct comparison with prior work. 
An alternative SPH formulations \citep{Ott2003} based on discretizing the particle 
number density instead of mass density, similarly to the density estimate approach in MFM, 
can also resolve the sharp core-mantle boundary. Recent tests using a similar scheme \citep{Solenthaler2008} found that it is difficult to build equilibrium models of planetary bodies. As a result this method might not be suitable for planetary-size collisions (Alexandre Emsenhuber, private communication).

We note that we use different mass particles for iron and dunite but that these masses are the same in both the impactor and target. Using different iron/dunite particle masses in the impactor and target can lead to numerical differentiation and thus cause unphysical mixing in our test runs with MFM.

\section{Results}
\subsection{Single component impact}
\label{sec:single}

For the single component models, we use the Tillotson EOS  because it is simple and highly reliable. This EOS can accurately model shocks, which are very important in high-speed impacts, 
and shows good agreement to measured data \citep{Brundage2013}. Its main weakness is that it does not provide a thermodynamically consistent treatment of vaporization, which is not an issue in this simulation as we mainly focus on the different inner structure of the post-impact target here.

We use 500K particles to represent a $1M_{\oplus}$ target and a $0.1M_{\oplus}$ impactor, both of which are composed of granite described by the Tillotson EOS. This one component model is free of core-mantle discontinuity which is hard to handle in numeric models, see the discussion in section \ref{sec:coremantle}. The impact setup is similar to the canonical Moon formation impact of \citet{Canup2013}. The impact velocity equals $10$ km/s (1.1 times of their mutual escape velocity) and the impact parameter $b=0.71$ corresponds to an impact angle of $45^{\circ}$. The initial separation between the two bodies is $0.4R_{\oplus}$. We run this simple impact with both 
standard SPH and MFM implementations in the GIZMO code, hence the only difference is the hydro-method.

We observe a striking difference in the inner structure of the post-impact target between MFM and standard SPH.
In figure \ref{fig:ids}, we mark three layers of the pre-impact target and the impactor with four different colors 
to trace the deformation of the target and the spread of the impactor. In the SPH simulation, the target's center deforms 
slightly while in the MFM simulation the center is dispersed throughout the body. 
In the SPH simulation, the outermost layer is strongly deformed but never penetrates the central region. The MFM method, instead,  allows fluid elements from the outermost layer of the target to mix  into the innermost region.
\begin{figure}[ht!]
\plotone{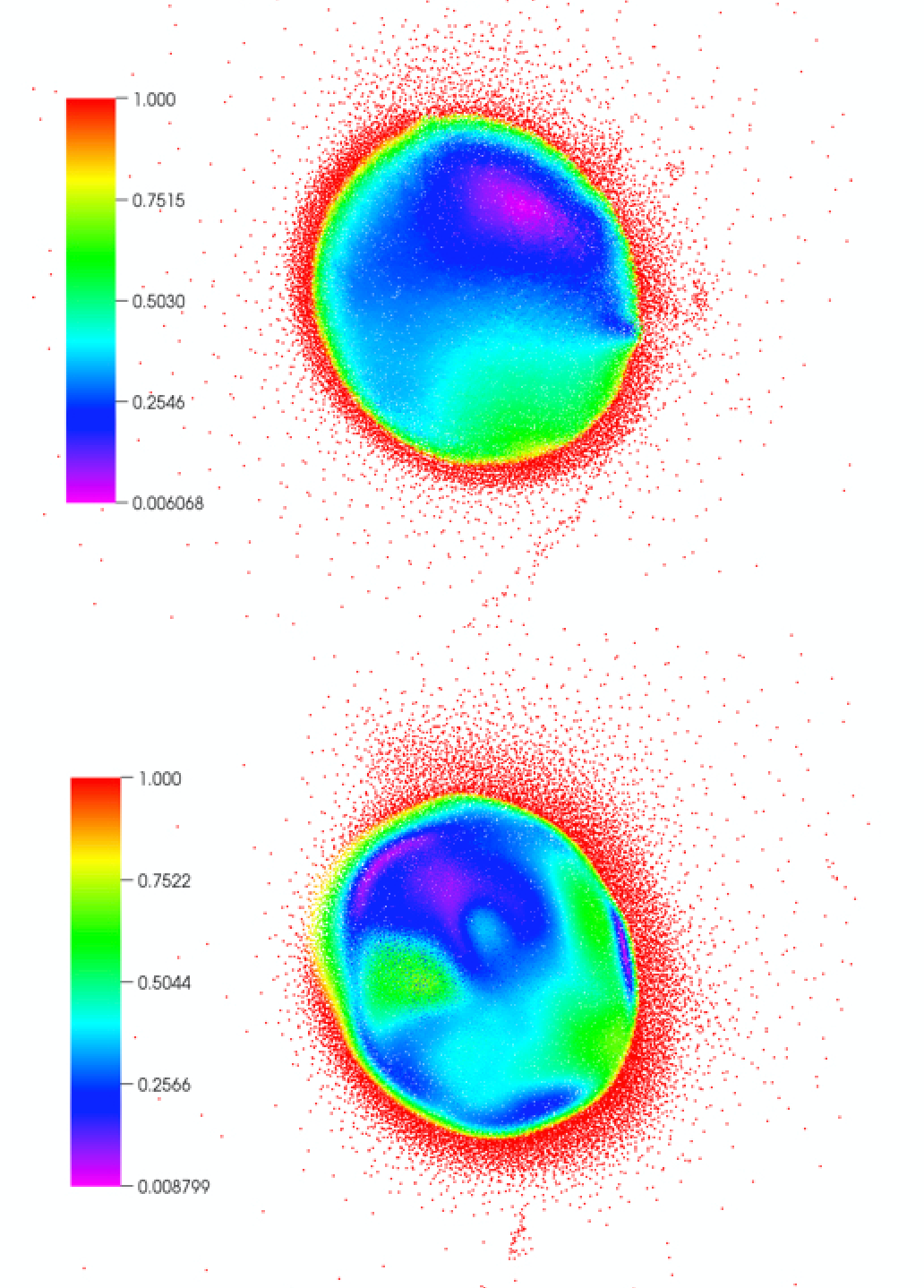}
\caption{The velocity magnitude of the $-0.1<z<0.1$ region in the major body of the single component impact \ref{sec:single}. The snapshots are taken at $t=10.5\,h$ and some clumps are still re-colliding with the major body. The upper panel is the SPH simulation and the lower panel is MFM simulation. MFM is able to capture the more complex subsonic 
turbulence while SPH tends to damp it readily on large scales, resulting in a more coherent flow rotating around a low-velocity center. \label{fig:vel}}
\end{figure}

This mixing happens  as a result of complex 3D subsonic turbulence whose characteristic velocity amplitude is
less than $1$km/s. Figure \ref{fig:vel} shows the velocity field around the $z=0$ plane after the giant impact.
In the SPH run, the flow is almost laminar and simply 
circulates around a low-velocity center. 
The flow structure is influenced by the tidal force from the ejecta and by their fall-back. 
In the MFM run, we always observe significantly more substructure in the flow characterizing the post-impact 
target. Our findings echo the analysis carried out by
\citet{Bauer2012}, who showed that standard SPH result
in a dissipation scale for turbulence that is unphysical and much higher than that
of finite volume methods using static or moving meshes. 
We expect behaviour of MFM
in this domain to be closer to the latter codes than to SPH as a result of the absence of 
explicit numerical dissipation from artificial viscosity and because of higher accuracy of velocity
variations computed by means of the Riemann solver. It is aligned with the outcome of the many numerical tests discussed in 
\citet{Hopkins2015}.
In the following section,  we  will assess the importance of capturing mixing promoted by (subsonic) turbulence 
in the  context of the canonical Moon-forming impact.

\subsection{Multiple-component impact}
\label{sec:multiple}
We simulated ``run119'' described by \citet{Canup2013} with SPH and MFM using 500K particles. In this impact, 
a $0.89M_{\oplus}$ target is hit by a $0.13M_{\oplus}$ impactor at their mutual escape velocity $\sim 9$km/s. This model was proposed as a benchmark by \citet{Barr2016} (see section 2.1). M-ANEOS coefficients and other details about the simulation setup may be found there. We note that we choose this model just because the Moon-forming impact is well studied. We are not trying to solve the isotope conundrum in the giant impact hypothesis of the Moon formation \citep{Asphaug2014} here but focusing on the different mixing in general impacts caused by the hydro-method. We also vary the impact velocity and angle and summarize our simulations in table \ref{tab:simulations}.

\begin{table}
\small
\caption{Comparison between SPH and MFM simulations\label{tab:simulations}}
\medskip
\begin{tabular}{@{}llllllll}
\hline
\hline
Run  & $b$ & $\frac{v_{imp}}{v_{esc}}$  & $\frac{L_D}{L_{EM}}$ & $\frac{M_{D}}{M_L}$ & $F_{D,tar}$ & $\frac{M_{Fe}}{M_D}$ & $\delta f_T$\\
\hline
  1         & 0.72&  1.0                       &0.35                  & 1.70                 & 0.27         & 0.07   & -0.70\\
  2         & 0.64&  1.0                       &0.05                  & 0.28                 & 0.47         & 0.10   &-0.48 \\
  3         & 0.64&  1.1                       &0.12                  & 0.63                 & 0.45         & 0.16   & -0.50 \\
  4         & 0.72&  1.0                       &0.35                  & 1.72                 & 0.30         & 0.07   & -0.67 \\
\hline
  5         & 0.72&  1.0                       &0.37                  & 1.86                 & 0.43         & 0.04   & -0.50 \\
  6         & 0.64&  1.0                       &0.06                  & 0.43                 & 0.82         & 0.04   & -0.08\\
  7         & 0.64&  1.1                       &0.12                  & 0.71                 & 0.62         & 0.15   & -0.30 \\
\hline
\hline
\end{tabular}
Note. Runs 1-4 are SPH simulations while runs 5-7 are MFM simulations. Runs 1-3 use equal mass rock/iron particles as in most previous studies while runs 4-7 use rock/iron particles of 1:2 mass ratio (see the discussion in section \ref{sec:coremantle}).
\end{table}

\subsubsection{Protolunar disk property}
We carry out the analysis of the protolunar disk following \citet{Canup2013}. We calculate the disk mass $M_D$ and disk angular momentum 
$L_D$ at $t=35$ h, when the properties of the disk no longer change significantly. 
In our SPH simulation (run1), we get a disk mass $M_{D}=1.70 M_{L}$ and disk angular momentum $L_{D}=0.35L_{EM}$, which is close to the results of the highest resolution simulation for run119, $M_{D}=1.69M_{L}$ and $L_{D}=0.33L_{EM}$ \citep{Canup2013}. 
 Here, $M_{L}$ and $L_{EM}$ are, respectively, the Moon mass and the angular momentum of the Earth-Moon system. 
Our SPH simulation agrees very well with previous SPH simulations. By comparing run4 which run1, which use particles with different masses in SPH, we conclude that changing the mass of particles does not make a significant difference. In our MFM simulation (run5), we have $M_{D}=1.86M_{L}, 
L_{D}=0.37L_{EM}$. Comparing runs 5-7 to 1-3, MFM simulations have larger disk mass and angular momentum than their SPH counterparts, which we
attribute to more accurate handling of angular momentum transport in MFM for differentially rotating flows \citep{Deng2017}
 We note that, while SPH  conserves angular momentum by construction, the inclusion of artificial viscosity causes dissipation that enhances angular momentum transport.

\subsection{Mixing}
\label{sec:mixing}

\begin{figure*}[ht!]
\plotone{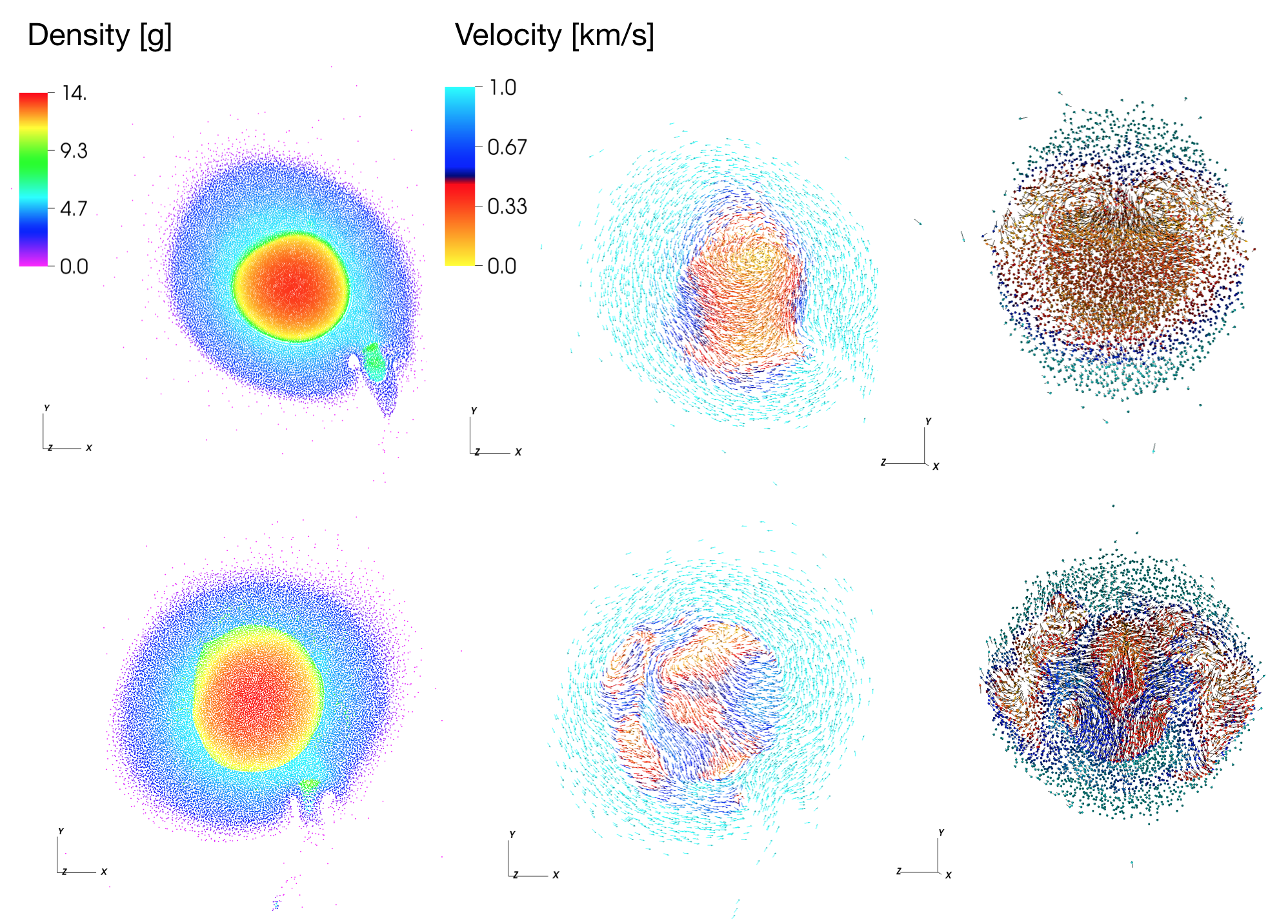}
\caption{Snapshots of multiple-component impact run 3 (SPH, three upper panels) and run 7 (MFM, three lower panels) at $\sim$7h.  Left two panels, density plot (slice between $-0.05<z<0.05$) of the post-impact target. SPH artificial tension force causes numeric particle separation which is absent in the MFM simulation. The rest four panels show the flow structure in the post-impact target in the x-y plane (Middle, impact plane) and y-z plane (Right). MFM captures much more complex three dimensional subsonic turbulence than SPH which is crucial to follow the mixing during the whole simulation time.\label{fig:vel2}}
\end{figure*}

In the canonical Moon formation scenario, the portion of the impactor that avoids colliding with the proto-Earth is sheared into spiral ejecta. The ejecta will contract and re-collide with the target and lead to the tidal 
disruption of the former and the formation of the disk. In this model, most of the disk matter comes from the tidal disruption of the impactor. In run119 of \citet{Canup2013}, 70\% of the disk material originates from the impactor.


Following \citet{Reufer2012} we use the deviation factor $\delta f_{T}$ to characterize the mixing in the Moon-forming giant impact,  where
\begin{align}
  f_{T}&=(M^{silc}_{targ}/M^{silc}_{tot})_{disk},  \\
  \delta f_{T}&=\frac{(M^{silc}_{targ}/M^{silc}_{tot})_{disk}}{(M^{silc}_{targ}/M^{silc}_{tot})_{post-impact  Target}} -1.
\end{align}
$M^{slic}_{targ}$ and $M^{slic}_{tot}$ denote the mass of the silicate part of the disk/post-impact target derived from the target and the total disk/post-impact target mass, respectively. $\delta f_{T}$ measures the composition similarity between the silicate part of the proto-lunar disk and the post-impact target.  In our SPH simulation, $f_{T}=27\%$, $\delta f_{T}=-70\%$ agrees well with $f_{T}\approx 30\%$ in \citet{Reufer2012, Canup2013}. In the MFM simulation, $f_{T}=43\%$, $\delta f_{T}=-50\%$ and there is a higher degree of mixing. This trend holds when we vary the impact angle and velocity (see table \ref{tab:simulations}).

\begin{figure*}[ht!]
\plotone{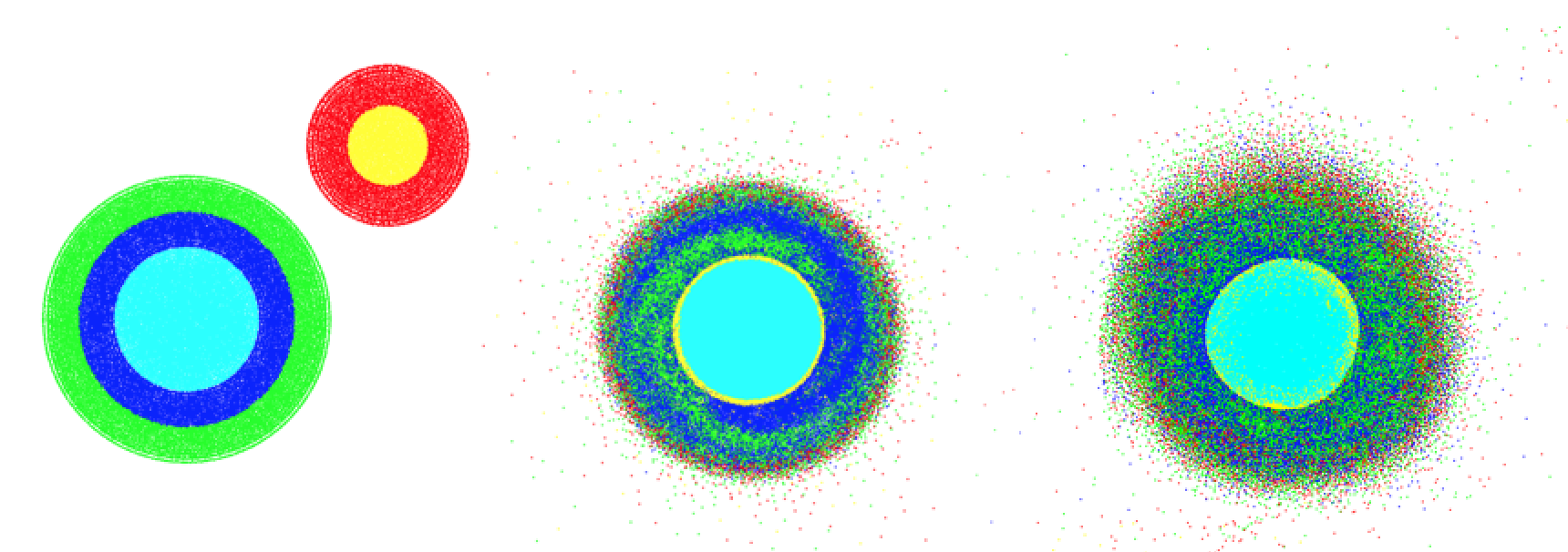}
\caption{Multiple-component impact with ANEOS/M-ANEOS. Left panel, color-labeled different layers (slice between $-0.1<z<0.1$) of the pre-impact target (core and two layers of mantle) and the impactor (core and mantle). Middle panel, the material distribution at $t=36\,h$ in the SPH run. Right panel, the material distribution at $t=14\,h$ in the MFM run. In the SPH simulation, particles from the impactor mantle stay on the surface of the post-impact target due to the artificial tension force at the surface of the target and suppression of turbulence in the inner part of the target, which is also shown in \citet{Emsenhuber2017}. However, MFM mixes the post-impact target thoroughly and quickly. MFM has a puffy planet surface which is similar to the density independent SPH of \citet{Hosono2016}.\label{fig:ids2}}
\end{figure*}

Similar to the single component model (figure \ref{fig:vel}), MFM captures more complex turbulence in multiple components impact (see figure \ref{fig:vel2}). In Figure \ref{fig:ids2} we can clearly
appreciate how different is the mixing in the two methods. We label with different colors the two layers of the proto-Earth mantle, core and impactor's mantle and core to trace the components. In the SPH simulation, the two layers of the mantle are distorted and become intertwined but do not mix (see snapshot taken at $t=36h$). However, MFM mixes the two layers of the proto-Earth mantle and the impactor mantle thoroughly and quickly (snapshot taken at $t=14h$).  

The extent of mixing (see figure \ref{fig:ids2}) in the multi-component Moon formation simulation is much more pronounced than in the single component model in figure \ref{fig:ids}. The iron core can reflect pressure waves and shorten the sound crossing time scale in the post-impact target. This
facilitates mixing in the post-impact target. The tidal interaction between the core and mantle also drives turbulence and enhances mixing. In the SPH simulations, silicates from the impactor always stay on the surface of the post-impact target. They originate from fall-back ejecta. 
The artificial surface tension (see section \ref{sec:coremantle}) prevents them from entering the inner part of the post-impact target \citep{Hosono2016}, 
while the suppression of turbulence in the post-impact target (see Figure \ref{fig:vel2}) prevents them from mixing with the target further. These two numerical effects in SPH tend to increase the concentration of the impactor’s material at the surface layer of the target. Some fall-back clumps are able to accelerate fluid elements across the surface layer of the target, and then  launch them onto disk-like orbits. As a consequence,  in  the SPH simulation more impactor material, which should have mixed deeper into the target, can be ejected. On the other hand,
MFM mixes the impactor's mantle and the target quickly, hence more silicates from the target 
can be propelled into the proto-lunar disk.

\section{Discussion: variants of the SPH method}
\label{sec:discuss}

In the previous sections, we have shown how MFM can resolve subsonic turbulence and the associated mixing
in GIs, which instead standard SPH cannot.
The artificial tension force of standard SPH prevents fluid mixing, which in turn prevents
fall-back ejecta from mixing with the post-impact target (section \ref{sec:mixing}). Artifacts
due to artificial surface tension can be alleviated in SPH by introducing  
a conductivity term in the hydro equations \citep{Price2008, Read2012}, or by employing a 
more accurate integral-based 
gradient estimator \citep{Garcia2012, Rosswog2015}. We tested the former improvement. We reran the SPH simulation of run119  with artificial conductivity as suggested by \citet{Read2012}. Mixing in the post-impact target is marginally improved, 
with the impactor's mantle penetrating a little deeper and the two layers of the target's mantle fracturing after
a strong distortion rather than remaining intact as in standard SPH (see figure \ref{fig:ids2}). However, this run also results in iron particles floating on the post-impact target's surface, which is likely caused by the complex EOS. (see appendix A of \citet{Saitoh2016})

 Concerning other improvements that we did not test, it should be recalled that, since mixing 
is aided by the development of sub-sonic turbulence triggered by the collision, the ability to capture
the latter phenomenon should be considered as a requirement for any SPH variant to be capable of modeling
the correct physical behaviour in giant impacts. This is additional to removing artificial surface tension.
In this respect \citet{Hopkins2015} showed that DISPH does not help to sustain subsonic turbulence, although \citet{Wadsley2017} found
considerable benefits when a similar approach is combined with higher order kernels and a turbulent diffusion term.
\citet{Beck2016} shows their improved Cullen \& Dehen switch helps to 
sustain subsonic turbulence. We also rerun the same simulation with the Cullen \& Dehnen artificial viscosity prescription but
did not find any noticeable difference in the mixing. In summary, so far we could not determine if there is any combination of the
many proposed improvements to standard SPH that can capture turbulence and mixing in the context of giant impacts, which
MFM can instead do by design.

\section{Conclusions and perspectives}
\label{sec:conclusions}
We employed both SPH and, for the first time, a new Lagrangian method (MFM) to carry out GI simulations.  Our goal was to compare their
outcomes and determine if the degree of mixing depends on the numerical technique.
In our single component model with the Tillotson EOS, we find that turbulence, and thus mixing, is suppressed in the SPH simulation. We 
then simulated the canonical Moon formation model with the M-ANEOS EOS. Our MFM initial conditions accurately model the core-mantle boundary 
with no particles entering an unphysical state. Our SPH results are consistent with previous results reported in the literature. The MFM 
simulations agree well with SPH simulations in terms of disk mass and angular momentum but show an marked increase 
in the mixing between the impactor and the target. 
 
MFM is a well-established hydrodynamics method with no numerical features that would
exaggerate the mixing seen in these simulations. Instead, the implication from our work is that previous
simulations have under-estimated the amount of mixing that happens in real impacts, which is line with
notorious problems of standard SPH in capturing mixing in other astrophysical applications \citep{Agertz2007,Wadsley2017}.
Yet, the outcomes for the canonical Moon-forming impact obtained here still have disks originating primarily from
the impactor.
Fully resolving the isotope conundrum arising in the Moon formation giant impact
theory \citep{Asphaug2014} likely requires different initial conditions for the encounter.
Hit-and-run models, for example, those in \citet{Reufer2012}, could potentially result in a more
efficient mixing, provided enough material is launched into orbit to create a satellite of lunar mass. Based on our results, MFM would seem to be an ideal method
to pursue further studies of mixing under a variety of initial conditions of GIs. 
This work simply represents the first step in this direction.

\bigskip

We thank Stephan Rosswog, Philip Hopkins, Romain Teyssier and Martin Jutzi for useful discussions and James Wadsley for careful and useful comments on the first version of this paper which helped to improve it considerably. We thank the referee, Alexandre Emsenhuber, and the other anonymous referee for suggestions that improved the paper.
We acknowledge support from the Swiss National 
Science Foundation via the National Center for Competence in Research (NCCR) PlanetS. C.R. acknowledges support from SNF Grant in ``Computational Astrophysics" (200020 162930/1). Author Barr acknowledges support from NASA Emerging Worlds grant NNX16AI29G.

\software{GIZMO code \citep{Hopkins2015}, ballic \citep{Reinhardt2017}, VisIt}

\bibliographystyle{aasjournal}
\bibliography{references}

\appendix

\section{Validation of the HLLC Riemann solver}
\label{sec:test1}
\begin{figure*}[ht!]
  \plotone{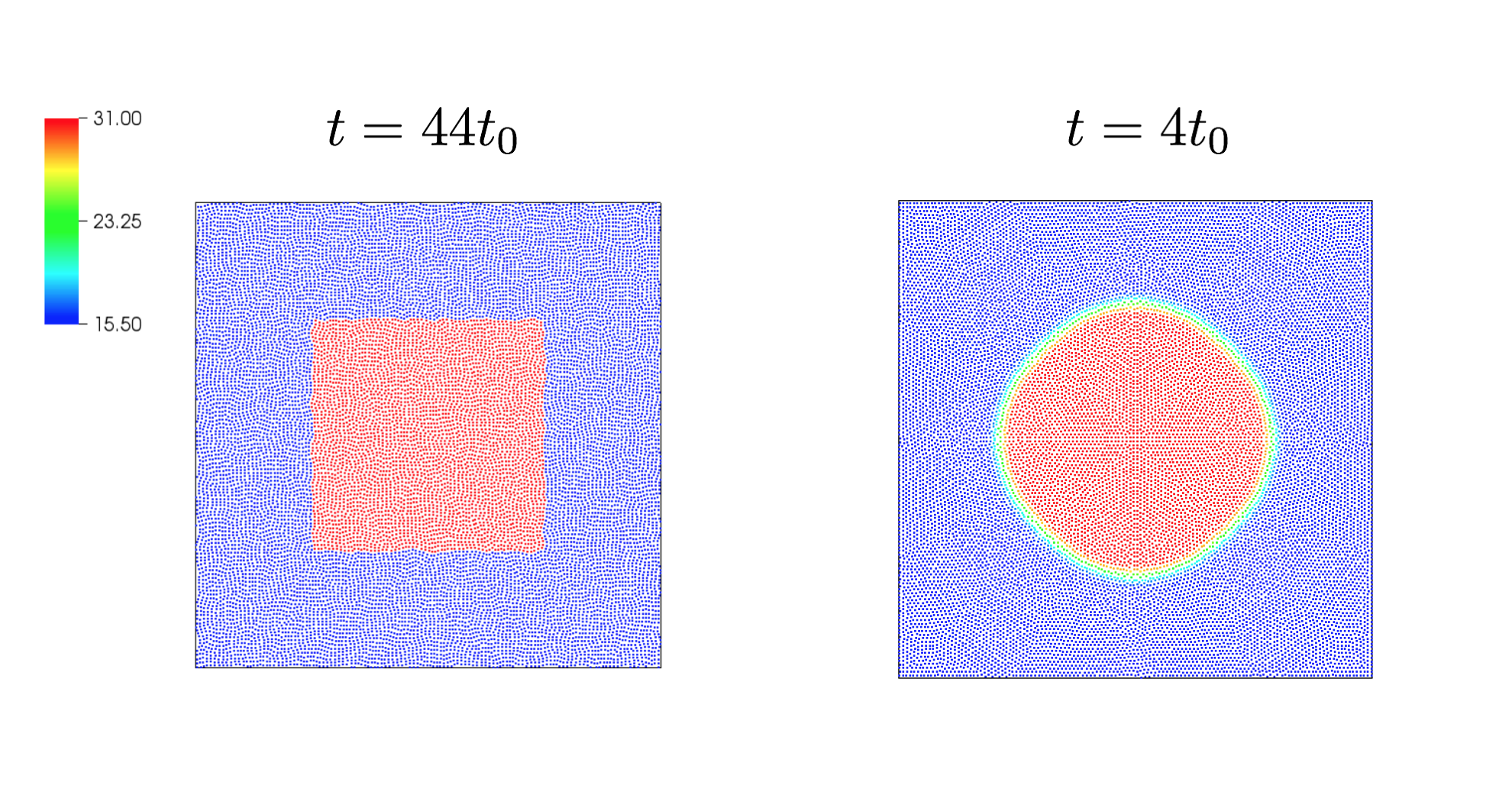}
  \caption{The density field in code units. Left panel: MFM solution maintains infinitely sharp density contrast. Right panel: the square quickly deforms into a circle due to the artificial tension force in the standard SPH simulation.\label{fig:square}}
\end{figure*}

We run the hydrostatic square test in \citet{Saitoh2013,Hopkins2015} using general EOS to test the HLLC Riemann solver at sharp boundaries. We initialize a two-dimensional fluid in a periodic box of Length $L=1$ (resolved by 128 particles) and uniform pressure $P=557.3$ (all in code units). We set ANEOS iron with $\rho=31$ within a central square of side-length $L=0.5$ surrounded by ideal gas with $\rho=15.5, \gamma=1.4$. The particles are evenly distributed but the iron particles' masses are twice of those of the gas particles. The sharp density contrast is well maintained at 44 sound crossing time (for the gas) in the MFM simulation and we observe no signs of deformation (figure \ref{fig:square}). Standard SPH cannot handle the sharp interface.

We then collide two uniform granite (Tillotson EOS) slabs ($15\times 15 \times 8 R_{\oplus}$) with opposite velocities. Both MFM and SPH in the GIZMO code can recover the Rankine-Hugoniot jump conditions \citep[e.g.][]{Melosh1989} which shows the code's ability to correctly capture shocks \citep{Reinhardt2017}.

\section{Entropy changes due to phase transitions}
\label{sec:test2}
Phase transitions can happen in impacts \citep{Kraus2011,Kraus2015} so entropy conservation is not guaranteed in GIs even when there are no shocks. Pressure release melting might ensue when the target is unloaded from highly compressed equilibrium state by the impactor \citep{Asphaug2006}. In the following tests we show that MFM can model phase transitions giving similar results to the CTH code in impact simulations. However, SPH cannot model phase transitions properly. We note that all the tests are run in the multi-method GIZMO code and all the comparisons are done with everything fixed except the factor we are discussing.

\begin{figure*}[ht!]
  \centering
  \includegraphics[width=0.8\textwidth]{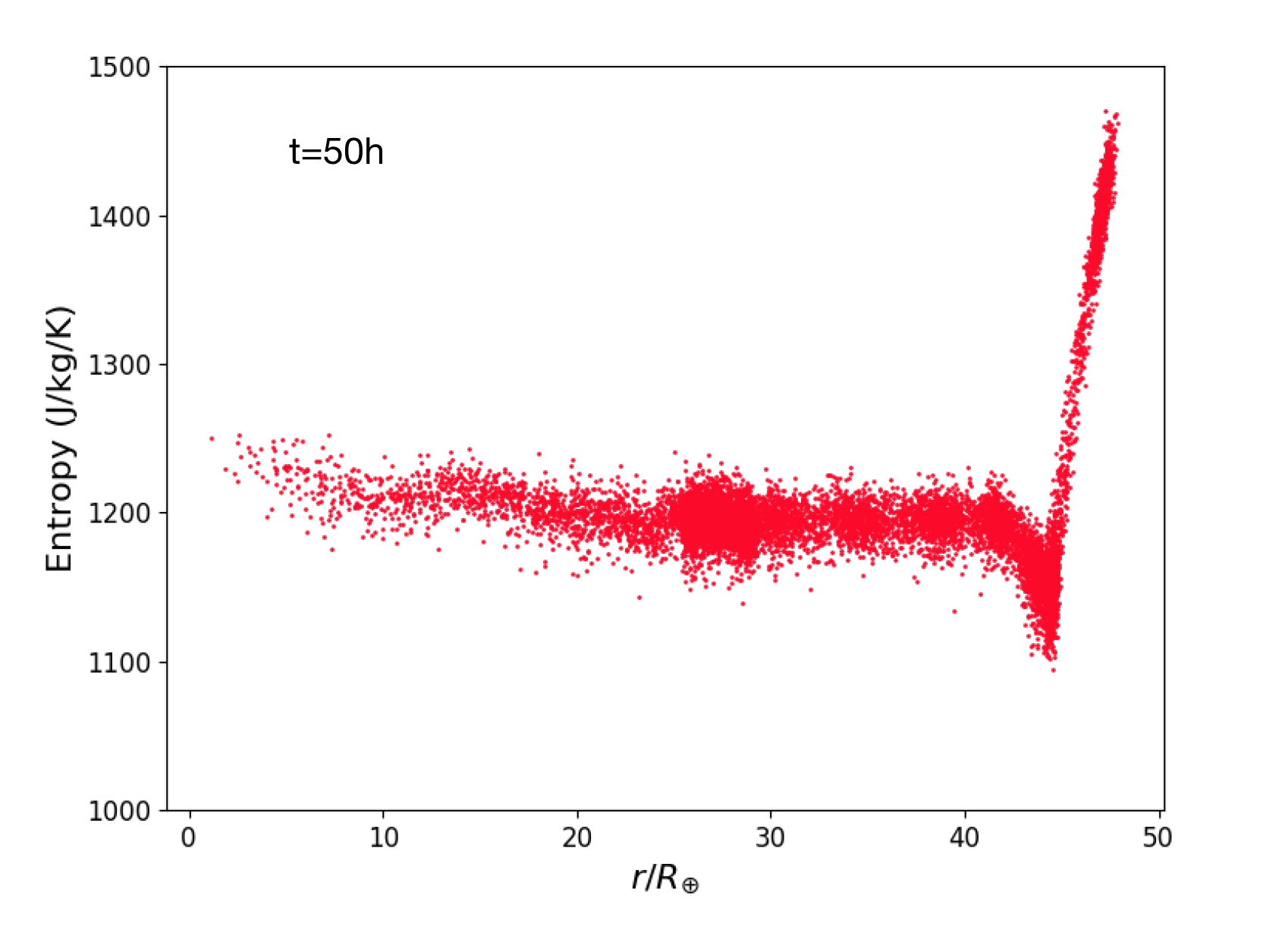}
  \caption{The entropy of the iron particles close to the equatorial plane ($−5 < z < 5$) in the MFM adiabatic expansion test.\label{fig:entr}}
\end{figure*}
\subsection{Conservation property}
We run two series of tests with the tabulated ANEOS EOS which has entropy information. The parameters for building the EOS table are set following \citet{Barr2016}. We take the iron core as an example to study the phase transitions.

First, our GIZMO code conserves entropy well both in SPH mode and MFM mode. Adiabatic expansion and pressure release melting is isentropic \citep{Pierazzo1997}. We run an adiabatic expansion test \citep{Reinhardt2017} by turning off the gravity of our $0.89M_{\oplus}$ target model in figure \ref{fig:2m} to test the entropy conservation of MFM. The iron core has an initial entropy of 1200J/kg/K. At 50 hours, the target expands about 100 times in radius; the resolution decreases a lot. The entropy of the iron particles are well conserved with deviation smaller than 3\% for most particles (see figure \ref{fig:entr}). At the core-mantle boundary, iron particles interact with the mantle leading to entropy non-conservation. MFM is able to conserve entropy well in the simulation time scale and the core-mantle boundary doesn't introduce systematic errors. SPH conserves entropy equally well in this test.

\begin{figure*}[ht!]
  \epsscale{0.8}
  \plotone{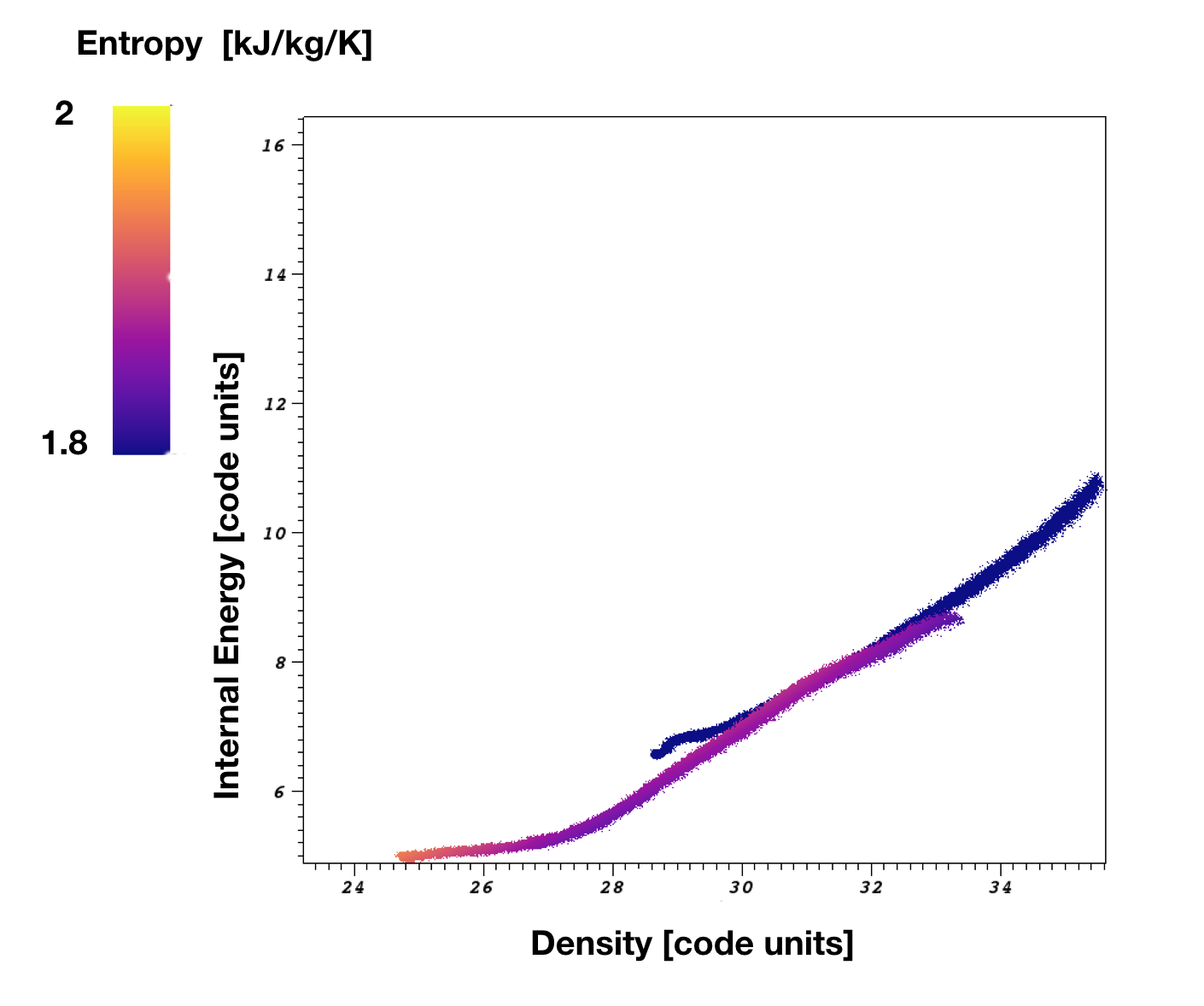}
  \caption{The phase diagram ($\rho -u$) of the fully molten iron core in the MFM oscillation test.  Particles lies on an isentrope with an entropy of 1860J/kg/K initially (dark blue particles). The iron core oscillates along the isentrope but the entropy remains $\sim$1860J/kg/K after 2.5 hours (see the color coded curve).\label{fig:phase0}}.
\end{figure*}

Second, our GIZMO code conserves the total energy well (internal energy plus kinetic energy and gravitational potential energy). We did an oscillation test on a hot $0.89M_{\oplus}$ protoplanet ($\sim$ 500 000 particles) by adding 1km/s radial velocities to particles beyond $0.7R_{\oplus}$. The surface temperature of the protoplanet is 10 000K and it has a fully molten core with an entropy of 1860J/kg/K \citep{Pierazzo1997}. It oscillates radially and the errors of the total energy are within 2\% in both MFM and SPH simulations. The core is fully molten and there are no phase transitions during the oscillations. The entropy of the core is well conserved shown in the phase diagram of figure \ref{fig:phase0}

\subsection{Phase transitions and internal energy redistribution}
\begin{figure*}[ht!]
  \plotone{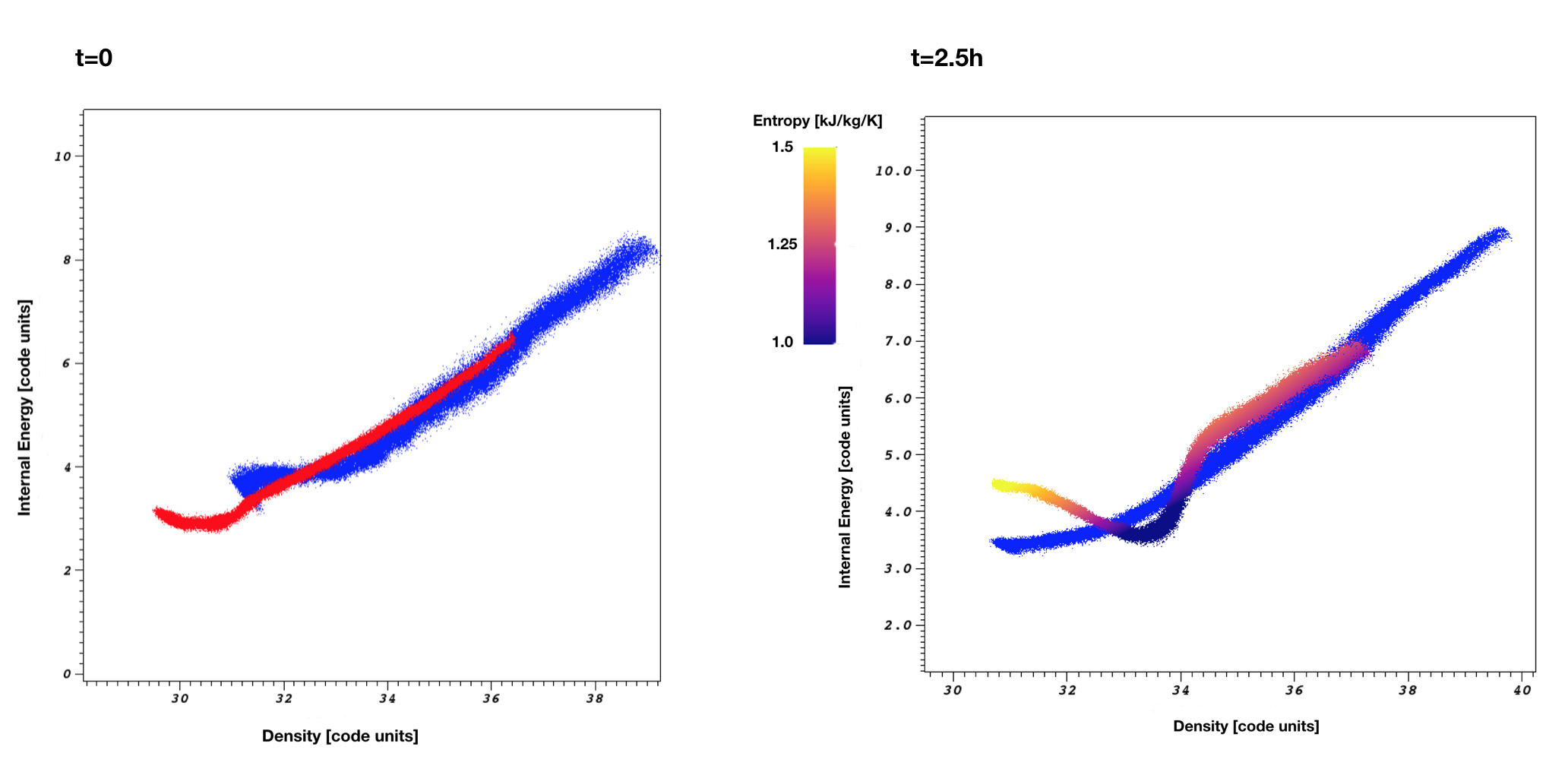}
  \caption{The phase diagram ($\rho -u$) of the iron core in the MFM oscillation test with the Tillotson EOS (blue particles) and ANEOS/M-ANEOS EOS (red or color coded by the entropy). Particles lies on an isentrope with an entropy of 1200J/kg/K initially, shown in the left panel. The iron core oscillates along the isentrope when we use the Tillotson EOS. In the simulation with the ANEOS/M-ANEOS EOS, the outer core melts due to pressure release and the internal energy of the core redistributes.\label{fig:phase1}}
\end{figure*}

In reality the earth doesn't have a fully molten core. In the moon formation impact simulations, the surface temperature of the proto-Earth is usually assumed to be $\sim$2000K and the core is close to the melting curve \citep{Alf1999,Anzellini2013,Barr2016}. When such a proto-Earth oscillates, pressure release melting starts at the outer core during the expansion; the outer core is more susceptible to melting than the inner core. Although pressure release melting is isentropic here it's not allowed to expand freely. The total energy flux ignoring the source term of the gravitational energy is $\bm{\nabla}(\rho u+\frac{1}{2}\rho v^{2}+P)$\citep{Hopkins2015}. The energy flux will soon reestablish quasi pressure equilibrium in the whole system. As a result, the melts near the CMB have higher internal energy than solid iron under the same pressure (see figure \ref{fig:phase1} right panel). During the compression, the high internal energy melts result in net energy flux to the mantle leading to thermal energy extraction from the central core. We will show this can be modeled with the ANEOS/M-ANEOS EOS in the following tests but not with the Tillotson EOS.

The Tillotson EOS lacks thermodynamically consistent treatment of mixtures between two phases and thus cannot model phase transitions \citep{Brundage2013}. However, ANEOS can indeed model phase transitions \citep{Melosh2007}. We run the oscillation test on the $0.89M_{\oplus}$ target model (see figure \ref{fig:2m}) whose core is close to the melting curve.  We build another $0.89M_{\oplus}$ target model using the Tillotson EOS. The mantle is granite instead of dunite. The iron core (blue particles) is slight more compressed than that of the ANEOS/M-ANEOS model (red particles) but they lie on the same isentrope (1200J/kg/K) (see left panel of figure \ref{fig:phase1}). 

In the Tillotson EOS simulation, the \emph{entropy conservation} is good and the core oscillates along the isentrope. However, the outer core melts, according to \citet{Pierazzo1997,Barr2011}, in the ANEOS/M-ANEOS EOS simulation (see right panel of figure \ref{fig:phase1}). This entropy changes is not a numeric artifact because even the same test with a fully molten core conserves entropy precisely (see figure \ref{fig:phase0}). It is a sign of internal energy redistribution and phase transitions in the core.

\subsection{Comparison between hydro-methods}
We run the oscillation test on a $0.89M_{\oplus}$ protoplanet model (core entropy 1200J/kg/K) with SPH and ANEOS/M-ANEOS. The entropy of the core slightly increase due to numerical dissipation from the artificial viscosity \citep{Springel2005}. The center of the core is strongly heated because strong artificial viscosity is wrongly triggered by the convergence flow during the compression \citep{Cullen2010}. We note that the central core's entropy increases by 300J/kg/K which is much larger than the core entropy gain ($\sim$100J/kg/K) due to the primary shock (at the first contact) in our simulations in table \ref{tab:simulations}. It shows no sign of phase transitions at the outer core because the pressure blips (see figure \ref{fig:2m}) help to separate the core and mantle. This numerical separation impairs energy flux and keep the core adiabatic to some extent.

\begin{figure*}[ht!]
  \epsscale{0.8}
  \plotone{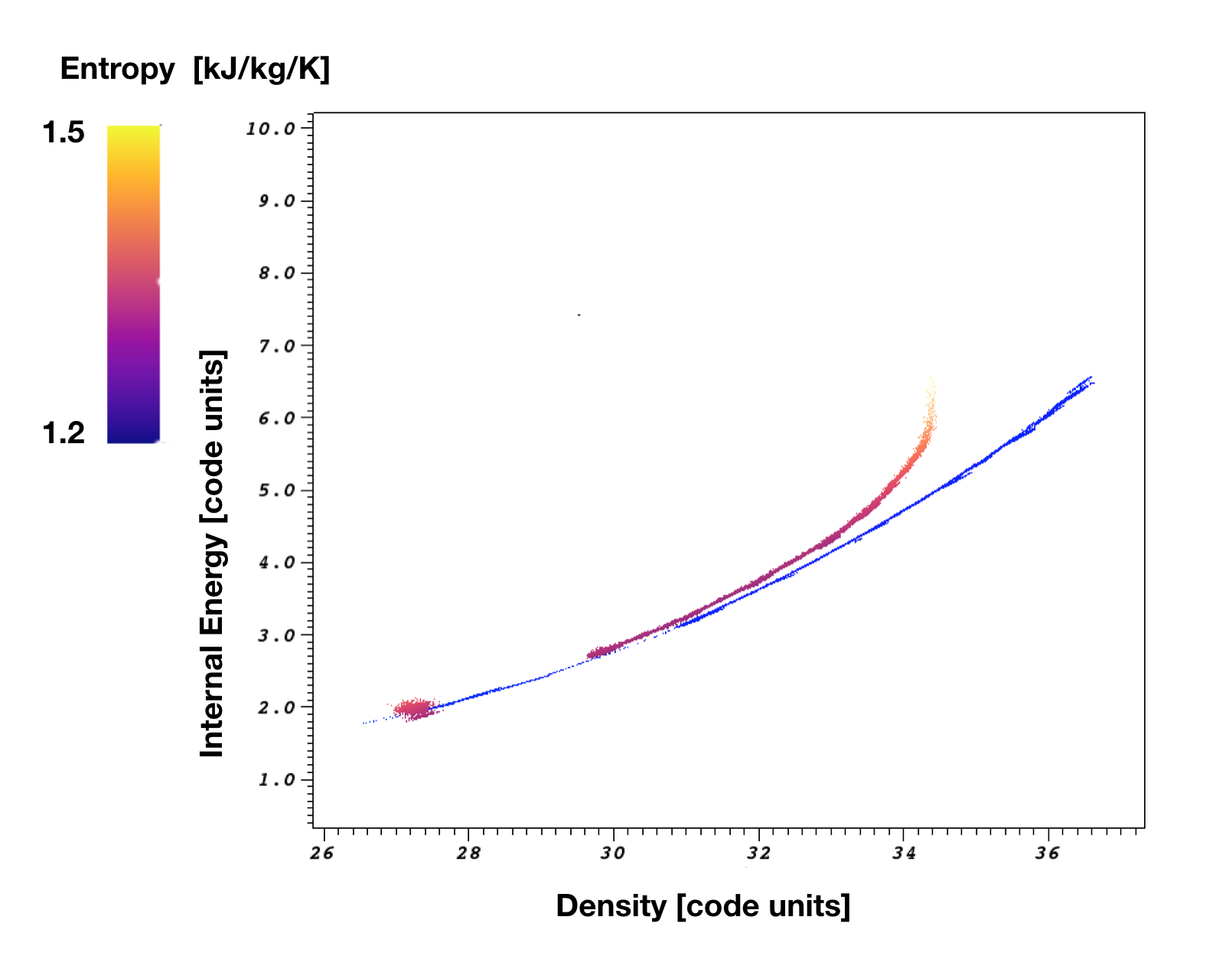}
  \caption{The phase diagram ($\rho -u$) of the iron core in the SPH oscillation test.  Particles lies on an isentrope with an entropy of 1200J/kg/K initially (blue particles). The iron core oscillates along the isentrope but the entropy increase $\sim$300J/kg/K in the central core after 2.5 hours (see the color coded curve).\label{fig:phase2}}
\end{figure*}

The CTH code is well tested with the ANEOS/M-ANEOS EOS\citep{Crawford2006}. Unfortunately we were not able to run the oscillation test with the CTH code. We checked the entropy profile of the post-impact target as a function of the normalized enclosed mass in the benchmark run119 \citep{Barr2016} by CTH and GIZMO using about 2M cells/particles.

\begin{figure*}[ht!]
  \plotone{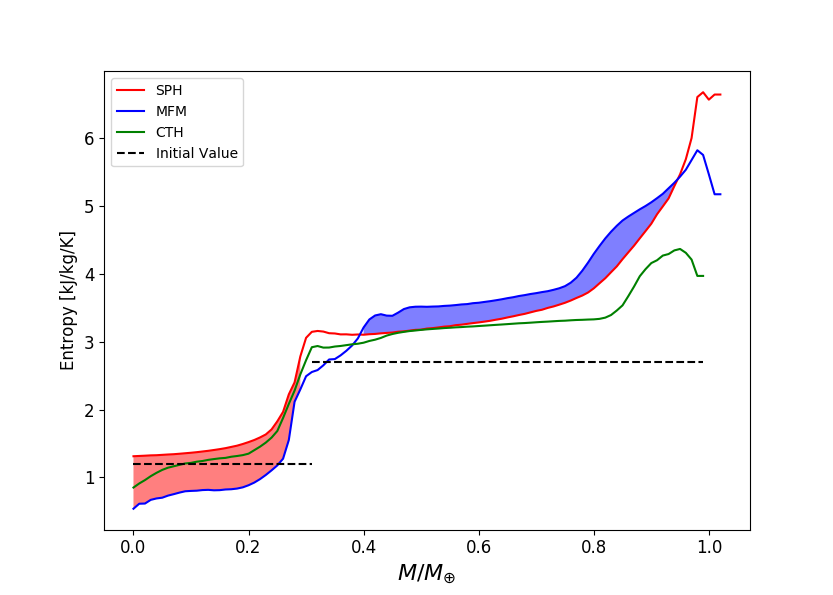}
  \caption{The entropy profile as a function of normalized enclosed mass in the CTH and GIZMO simulation ($t=40h$) of run119 with the initial entropy indicated by the black dash lines. The central core's entropy decreases in the CTH and GIZMO MFM simulation which is absent in the GIZMO SPH simulation. The difference in entropy around the core-mantle boundary is likely caused by the different treatment of material interfaces. \label{fig:phase3}}
\end{figure*}

First, the entropy structure of the mantle (in figure \ref{fig:phase3}) agrees well with \citet{Nakajima2015} when we run GIZMO in SPH mode validating again our EOS implementation. In figure \ref{fig:phase3}, parts of the post-impact target's core have even lower entropy than their initial values (1200J/kg/K, indicated by the black dash lines) in both GIZMO MFM and CTH. It is known well that the shocks deposit thermal energy and increase the entropy. The entropy decrease can be explained by phases transitions in the outer core and the following internal energy redistribution (lost to the mantle) as discussed above. The red and blue shaded region are of almost equal area and they measure the extra thermal energy transport from the core to the mantle in the MFM run comparing to the SPH run.

The two code are very different by nature because CTH is an AMR Eulerian code and GIZMO is a Lagrangian code. The treatment of material interface is also different. In GIZMO (both MFM and SPH mode) every computational element is either iron or rock but CTH allow cells with both rock and iron contribution. Given all the difference above, the qualitative agreement in the thermal structure of the post-impact target (see figure \ref{fig:phase3}) is satisfactory.

\end{document}